\documentclass[]{aa}
\usepackage[varg]{txfonts}
\usepackage{natbib}
\usepackage{graphicx}
\usepackage{caption}
\usepackage{subcaption}
\usepackage[draft]{hyperref}
\usepackage{soul}
\usepackage{booktabs}

\begin{document}

\title{Multi-source self-calibration: Unveiling the microJy population of compact radio sources}

\author{J. F.~Radcliffe\inst{*1,2,3}
\and M. A.~Garrett\inst{2,4} 
\and R. J.~Beswick\inst{1}
\and T. W. B.~Muxlow\inst{1}
\and P. D.~Barthel\inst{3}
\and A. T.~Deller\inst{2}
\and E.~Middelberg\inst{5}}

\institute{Jodrell Bank Centre for Astrophysics/e-MERLIN, The University of Manchester, M13 9PL, United Kingdom
\and ASTRON, the Netherlands Institute for Radio Astronomy, Postbus 2, 7990 AA, Dwingeloo, The Netherlands
\and Kapteyn Astronomical Institute, University of Groningen, 9747 AD Groningen, The Netherlands 
\and Leiden Observatory, Leiden University, PO Box 9513, 2300 RA Leiden, The Netherlands
\and Astronomisches Institut, Ruhr-Universit\"{a}t Bochum, Universit\"{a}tsstr. 150, 44801 Bochum, Germany}

\date{Received <date> / Accepted <date>}

\abstract{Very long baseline interferometry (VLBI) data are extremely sensitive to the phase stability of the VLBI array. This is especially important when we reach $\mu$Jy r.m.s. sensitivities. Calibration using standard phase-referencing techniques is often used to improve the phase stability of VLBI data, but the results are often not optimal. This is evident in blank fields that do not have in-beam calibrators.}{We present a calibration algorithm termed multi-source self-calibration (MSSC) which can be used after standard phase referencing on wide-field VLBI observations. This is tested on a 1.6 GHz wide-field VLBI data set of the Hubble Deep Field North and the Hubble Flanking Fields.}{MSSC uses multiple target sources that are detected in the field via standard phase referencing techniques and modifies the visibilities so that each data set approximates to a point source. These are combined to increase the signal to noise and permit self-calibration. In principle, this should allow residual phase changes caused by the troposphere and ionosphere to be corrected. By means of faceting, the technique can also be used for direction-dependent calibration.}{Phase corrections, derived using MSSC, were applied to a wide-field VLBI data set of the HDF-N, which comprises of 699 phase centres. MSSC was found to perform considerably better than standard phase referencing and single source self-calibration. All detected sources exhibited dramatic improvements in dynamic range.  Using MSSC, one source reached the detection threshold, taking the total detected sources to twenty. This means 60\% of these sources can now be imaged with uniform weighting, compared to just 45\% with standard phase referencing. In principle, this technique can be applied to any future VLBI observations. The Parseltongue code, which implements MSSC, has been released and made publicly available to the astronomical community (\url{https://github.com/jradcliffe5/multi\_self\_cal}).}{} %change when all other changes have been applied. 


\titlerunning{Multi-source self-calibration}
\authorrunning{J.F. Radcliffe et al.}

\maketitle

\section{Introduction}

With the expanded performance and capabilities of VLBI arrays, such as the European VLBI Network (EVN), r.m.s sensitivities of the order a few micro-Jansky are attainable in just a few hours. This allows compact sources with brightness temperatures of just $10^4$-$10^5$ K to be detected. Improvements in correlator capabilities have enabled the possibility of wide-field VLBI operations. Originally, data sets were correlated at a single phase centre with an ultra high temporal and frequency resolution to allow the entire primary beam to be mapped \citep[e.g.][]{2001A&A...366L...5G, chi2013deep}. However, these kinds of methods result in large data volumes and a degradation in image quality towards the edge of the primary beam.

In recent years, the introduction of software-based correlators has established the concept of `multiple simultaneous phase centre observing' \citep{deller2011difx,keimpema2015sfxc}. This method uses multiple phase centres with a coarser temporal and frequency resolution to produce a narrow field data set per phase centre. This method parallelises the correlation process and, as such, the correlation speed is now limited by the number of nodes in the correlator. These phase centres can be arranged to cover the entire primary beam \citep[e.g.][]{2015MNRAS.452...32R}.  As a result, the practical number of sources that can be detected and imaged in one observation has dramatically increased. These improvements have enabled the entire primary beam of a typical VLBI telescope to be completely mapped out to milliarcsecond resolutions and microJy sensitivities.

\begin{figure*}[!htb]
        \centering
        \begin{subfigure}{.5\textwidth}
                \centering
                \includegraphics[width=1\linewidth]{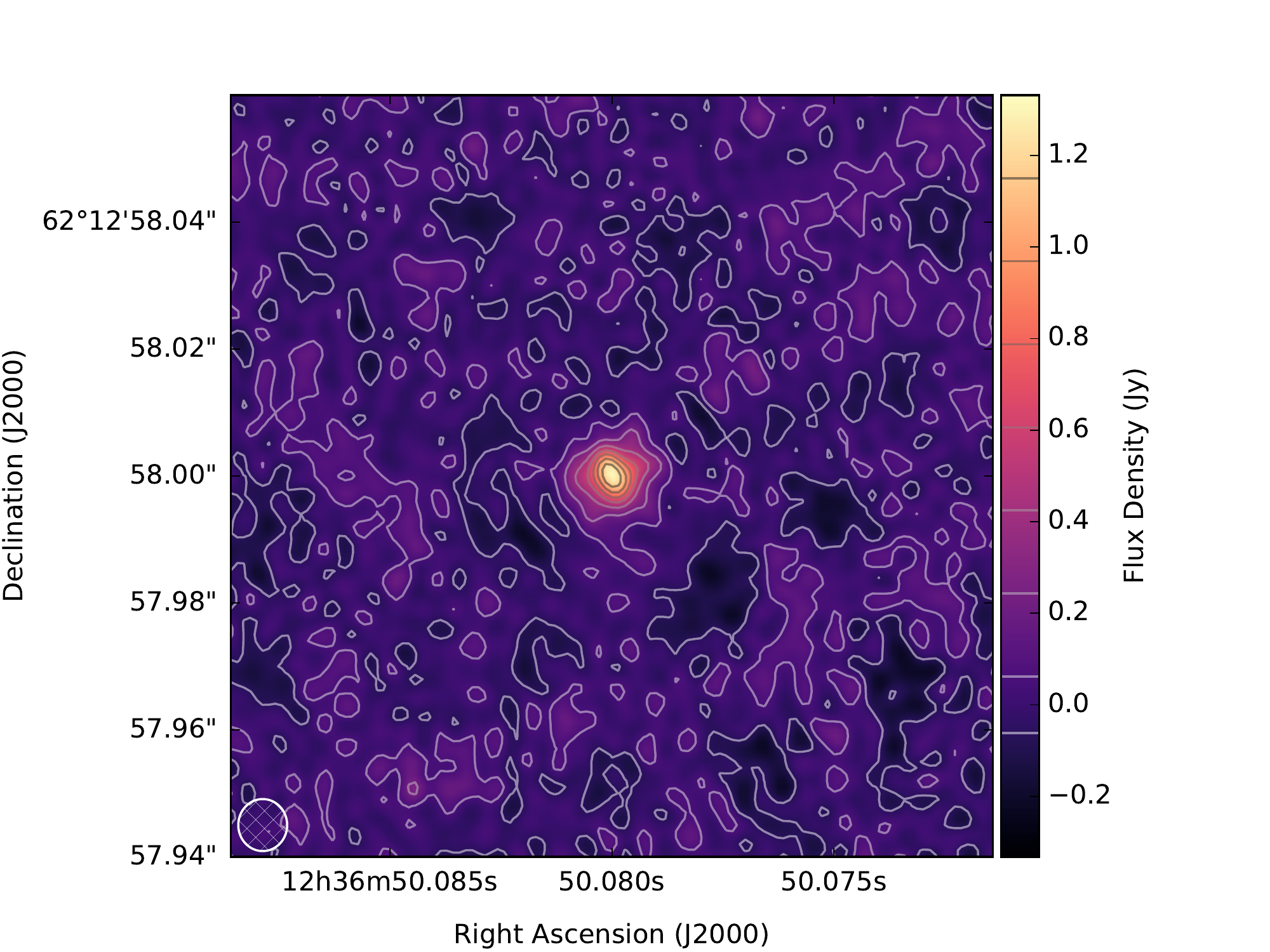}
                \label{fig:sub1}
        \end{subfigure}%
        \begin{subfigure}{.5\textwidth}
                \centering
                \includegraphics[width=1\linewidth]{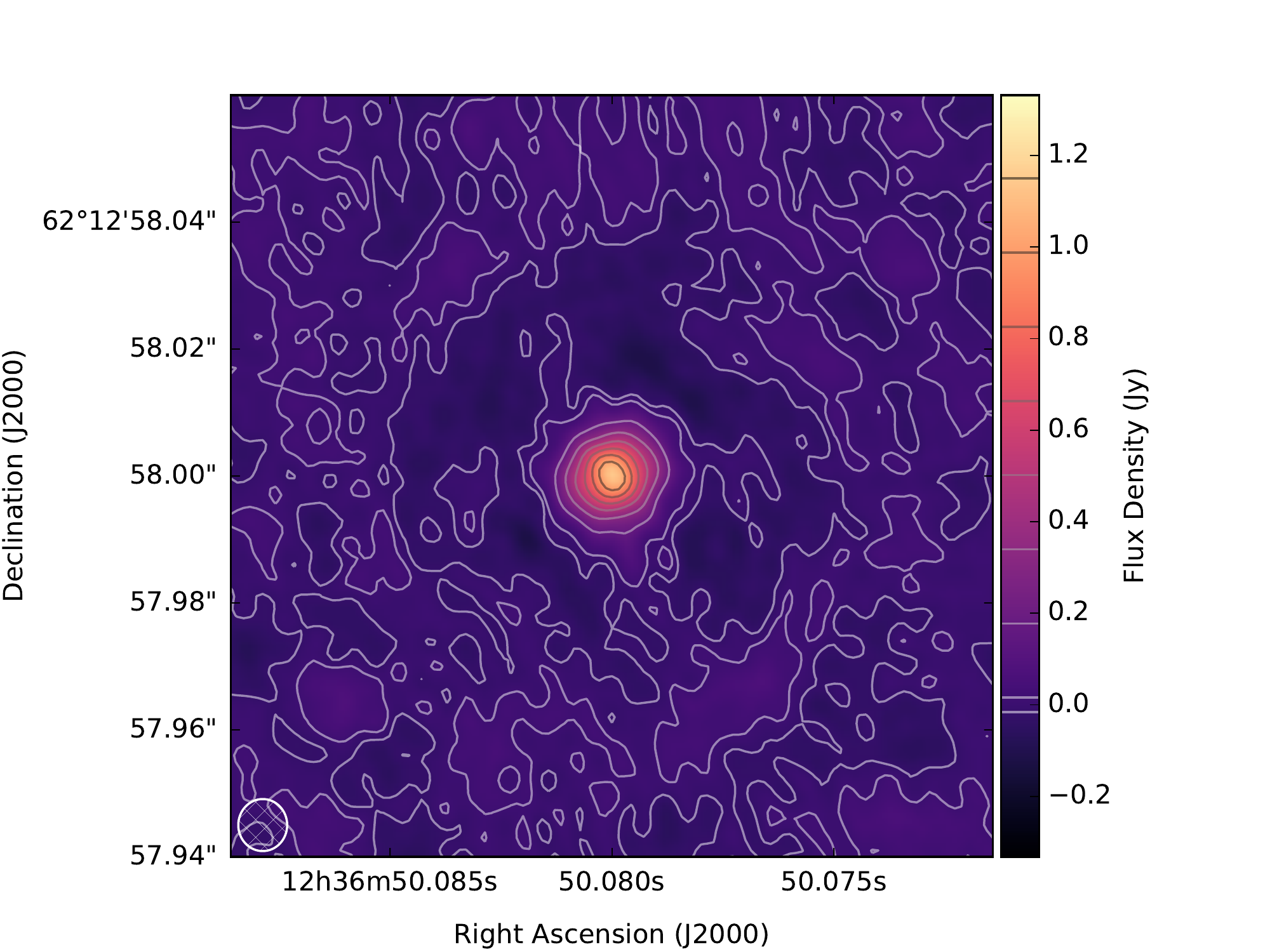}
                \label{fig:sub3}
        \end{subfigure}
        \caption{\textit{Left panel:} J123646+621405 when divided by the CLEAN model of the source. The low signal to noise results in an imperfect CLEAN model. This creates deviations from a normalised point source and, as a result, the peak brightness is $\sim1.4$ Jy/beam. Because the model cannot fully characterise the source structure, some of the flux density will be scattered into the side lobes. This results in a slightly reduced S/N of 7.7 compared to 7.9 before dividing the CLEAN model. \textit{Right panel:} Deconvolved image of 9 different, combined target sources, including the source in the left panel. Each source has been divided by its CLEAN model and combined to create a point source with a higher signal to noise. The deviations from a perfect, normalised point source have reduced and the peak brightness is $\sim1.1$ Jy/beam. This is the source used to self-calibrate the HDF-N data set in MSSC. The source morphology is more representative of a point source and the S/N has vastly increased to 93.1 which  allows self-calibration to be performed.}
        \label{fig:point_plots}
\end{figure*}

VLBI observations are particularly sensitive to the temporal and spatial variations of the troposphere and ionosphere. These cause phase variations over the course of an observation. To account for these, calibration on bright, nearby, and compact sources is essential. This is called `phase referencing'. It involves employing one or more compact sources nearby (or within the primary beam of) the target field to correct for gain and phase fluctuations. However, many target fields often do not  have compact sources, which can be used directly for calibration.

If this is the case, a chain of two or more sources can be used, which increase in brightness with respect to the distance from the target field. This `boot-strapping' approach allows phase calibration corrections to be derived from the brightest calibrator which is the furthest away and passed onto the next calibrator which is closer to the target. Self-calibration is used at each step to refine the corrections. This is repeated until the corrections derived for the nearest phase calibrator can then be applied to the target field. We note that amplitude calibration is only performed on the brighter calibrators with a sufficient signal to noise ratio (S/N). The phase corrections applied to the target field reduce in accuracy with respect to the angular separation between the target field and the final phase calibrator source because of the atmospheric inhomogeneities. If the angular separation is too large, the phase corrections derived are not fully representative of the atmosphere in front of the target field. As a result, the dynamic range of many VLBI targets can often be limited by phase errors. Accurate phase calibration becomes ever more important as the r.m.s. sensitivities continue towards the faint $\mu$Jy regime.
 
\begin{figure*}[!htb]
        \centering
        \begin{subfigure}{.5\textwidth}
                \centering
                \includegraphics[width=1\linewidth]{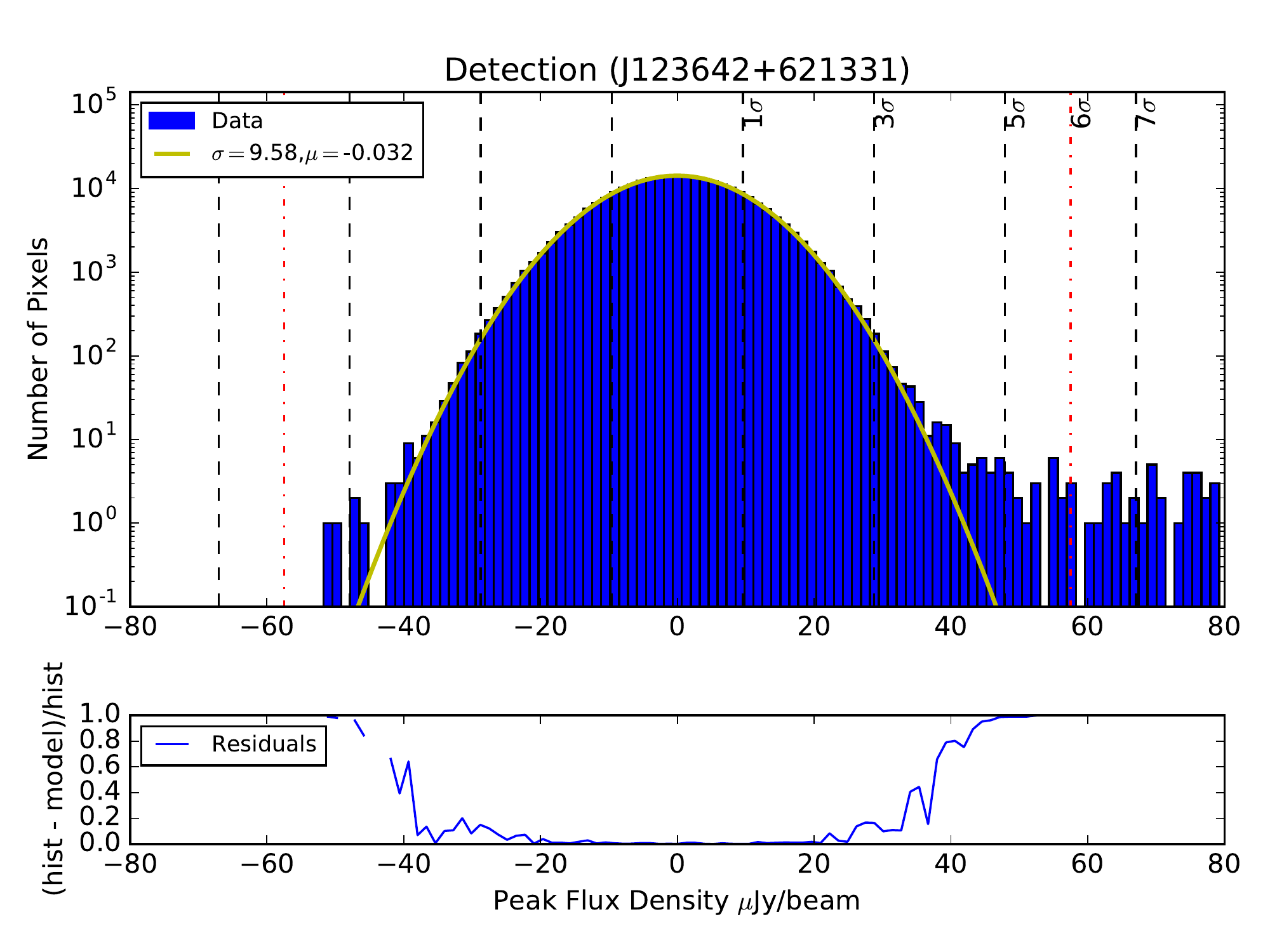}
                \label{fig:sub1}
        \end{subfigure}%
        \begin{subfigure}{.5\textwidth}
                \centering
                \includegraphics[width=1\linewidth]{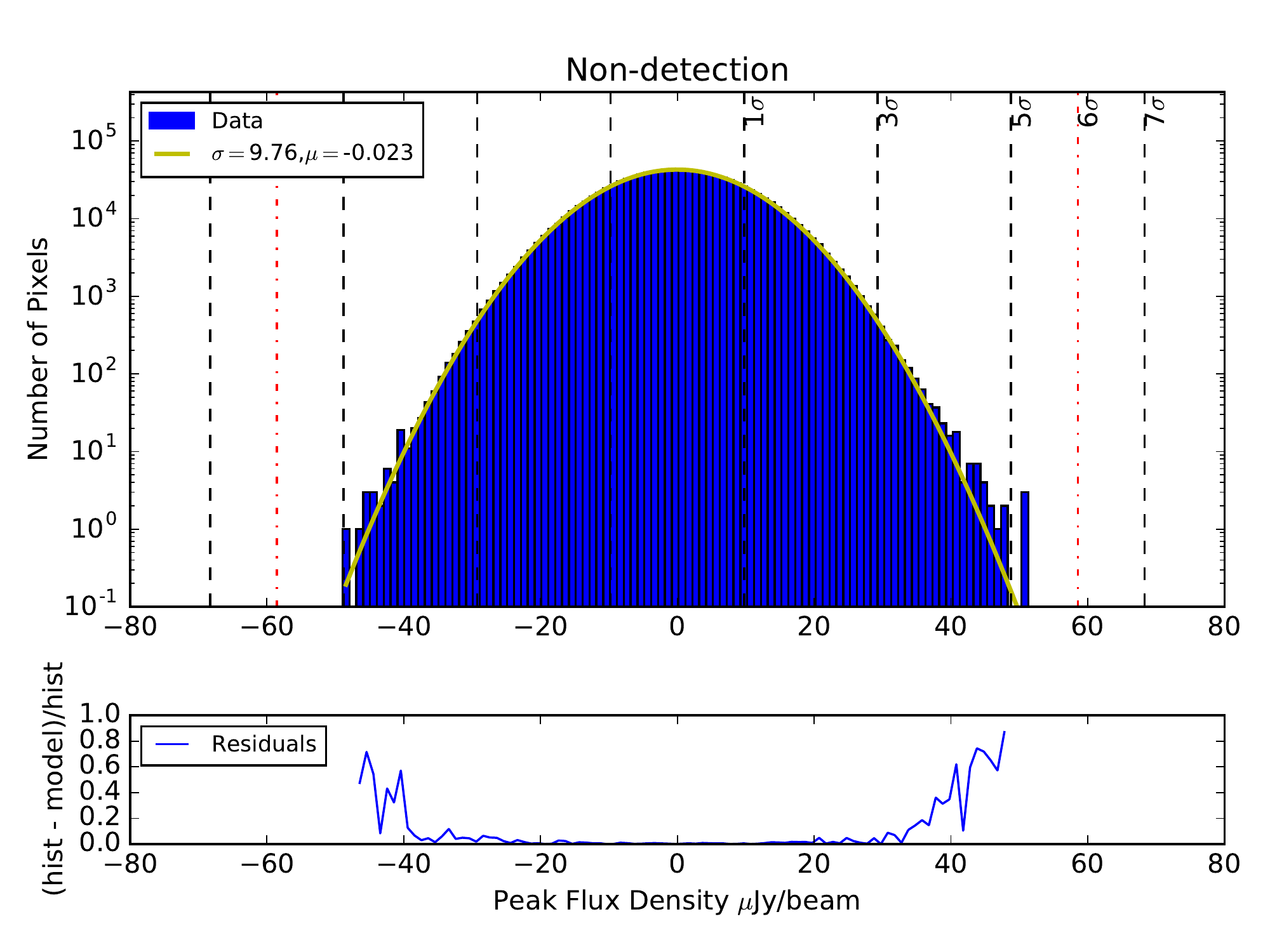}
                \label{fig:noise_distribution}
        \end{subfigure}
        \caption{Upper panels show histograms of the peak brightness distributions for fields with detected sources and no detected sources. Each histogram is derived from 1024x1024 pixel, uniformly weighted image and has a Gaussian distribution fitted to the noise profile. The vertical dashed lines indicate the values of $\pm1\sigma$, $\pm3\sigma$, $\pm5\sigma$ and $\pm7\sigma$ and the red dot-dashed line represents the $6\sigma$ detection threshold. The bottom panels show the residuals which are normalised to the histogram. \textit{Left panel:} Histogram of the field including the source J123642+621331. The fitted Gaussian approximates an r.m.s. noise level of 9.58 $\mu$Jy/beam. There are large deviations from the Gaussian model at the negative extrema of the flux distribution, which suggests that gain and phase errors are the cause. The deviations at the positive extrema is due to source structure. \textit{Right panel:} Histogram of a blank field with phase centre coordinates R.A. 12:36:05.0 and Dec. +62:12:30.0. A Gaussian with a $1\sigma$ r.m.s. noise level of 9.76 $\mu$Jy/beam is fitted. The residuals show smaller deviations from a Gaussian at the extrema of the flux distribution. This may be due to some residual radio frequency interference.}
        \label{fig:noise_distribution}
\end{figure*}
In principle, these errors can be corrected by performing self-calibration on the target field \citep{Trott2011}. Ordinarily, the response of a single, faint source is not sufficient to employ self-calibration. However, \citet{2000A&A...355..552R} and, in particular, \citet{2004evn..conf...35G} first demonstrated the potential of employing multiple sources detected across the primary beam as inputs for the self-calibration of large wide-field VLBI data sets. A previous VLBI survey of the Lockman Hole by \citet{2013A&A...551A..97M} developed the fundamentals of the technique presented here and has recently been employed in the analysis of VLBI observations of the Hubble Deep Field-North (HDF-N). 

Known as Multi-Source Self-Calibration (MSSC), this is a calibration technique that provides an additional step to standard phase referencing. MSSC is designed to be used for multiple phase centre correlated VLBI observations but, in principle, it can be used on any observation that targets multiple sources. MSSC uses multiple faint sources that are detected within the primary beam and combines them. The combined response of many sources across the field of view is generally more than sufficient to allow phase corrections to be derived. Each source has their CLEAN model divided into the visibilities\textbf{,} which results in multiple point sources. These are stacked in the UV plane to increase the S/N, which allows self-calibration to become feasible. The corrections derived can then be applied to the original phase-referenced data. It is worth noting that this process only applies to wide-field VLBI data sets that detect and image multiple sources within one epoch.  Recent improvements in the capabilities of VLBI correlators are ensuring that wide-field VLBI is a reality and, as a result, there will be an increased number of experiments which utilise MSSC. 

MSSC has been released and made publicly available to the astronomical community as a Parseltongue script \citep{kettenis2006parseltongue}. In this paper, we  demonstrate the power of this calibration technique upon one of the largest and most sensitive wide-field VLBI surveys ever conducted, which targets the HDF-N.

\section{Hubble Deep Field-North and Hubble Flanking Field wide-field VLBI observations}
\label{sec:2}
We have completed the first of three 24-hour epochs of a 1.6 GHz wide-field VLBI survey using the EVN array. The observations target a 15 arcminute diameter area centred on the HDF-N. This survey implements the `multiple simultaneous phase centre observing' mode of the SFXC correlator \citep{keimpema2015sfxc} to image a 7.5 arcminute radius area by simultaneously correlating on 582 phase centres. This enables us to achieve $\mu$Jy r.m.s. noise levels with milliarcsecond resolution across the whole of the primary beam. An additional 127 phase centres were used to target bright sources up to 12 arcminutes from the pointing centre. The total number of phase centres correlated is 699. The phase centres include 607 sources which were detected in the e-MERLIN eMERGE survey \citep{wrigley2014high} and the VLA \citep{morrison2010very}. 

Each phase centre produces a narrow-field (averaged) data set that can be calibrated in an identical fashion using the MSSC solutions. Compared to imaging the entire primary beam, this is both considerably less computer-intensive and much more easily parallelisable.  Since brute force surveying at VLBI resolution is computationally bound, this provides a way to greatly increase the effective (computationally feasible) survey speed of VLBI observations.

After standard phase referencing there were 19 detected sources, 18 of which are located in the central 7.5 arcminute radius area. The inner few arcminutes reach r.m.s. sensitivities of 5 $\mu$Jy/beam and this is expected to reach 1$\sigma$ thermal noise levels of $\sim$1.5-4 $\mu$Jy/beam (depending on telescope availability) with the addition of two further epochs. The scientific results of this survey will be presented in a future publication \citep{radcliffe2015AGN}. This represents a substantial improvement when compared with the previous VLBI observations of the field which had a central r.m.s. sensitivity of 7.3 $\mu$Jy/beam, presented by \citet{chi2013deep}.

\section{Multi-source self-calibration}
The HDF-N field was an ideal candidate for MSSC. This is a field with few bright sources in all wavebands. In the radio, the brightest sources have integrated flux densities of the order of a few mJy. Before MSSC can be utilised, phase referencing has to be conducted. All calibration steps were conducted using the Astronomical Image Processing System (AIPS) and its Python interface, Parseltongue \citep{kettenis2006parseltongue}. In the HDF-N data, this is comprised of two sources, J1241+602, a bright 0.4 Jy source located $2^{\circ}$ away from the pointing centre, and J1234+619, a faint 20 mJy source located 23.5 arcminutes away. The brighter calibrator was used to obtain phase and gain corrections. The solutions obtained were then applied to the fainter source. Further rounds of self-calibration were conducted to refine the phase corrections and then the calibration was applied to the target field. We note that there was insufficient S/N to perform gain calibration on this source. 
\begin{figure*}[!htb]
        \centering
        \begin{subfigure}{.5\textwidth}
                \centering
                \includegraphics[width=1\linewidth]{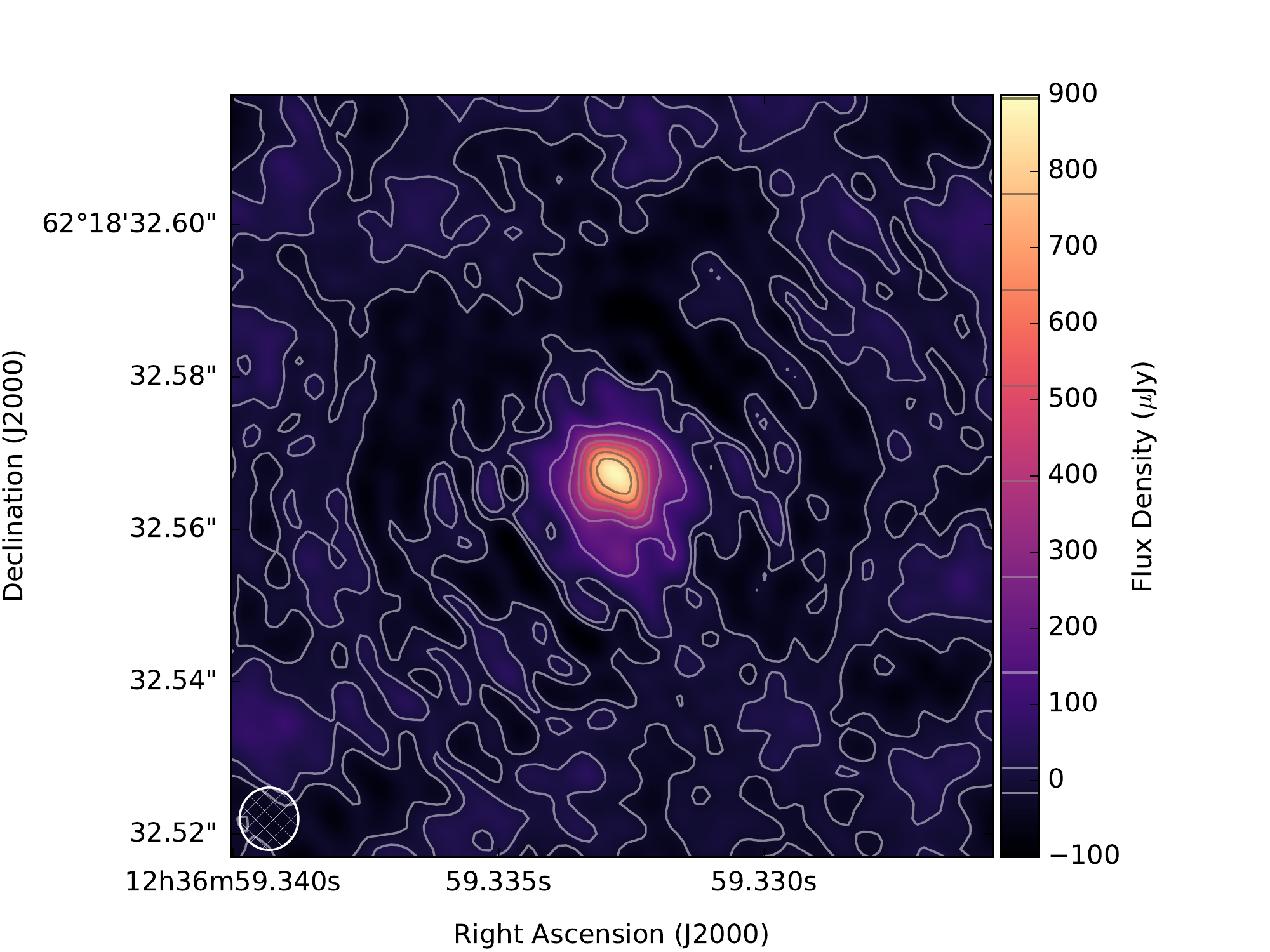}
                \label{fig:sub1}
        \end{subfigure}%
        \begin{subfigure}{.5\textwidth}
                \centering
                \includegraphics[width=1\linewidth]{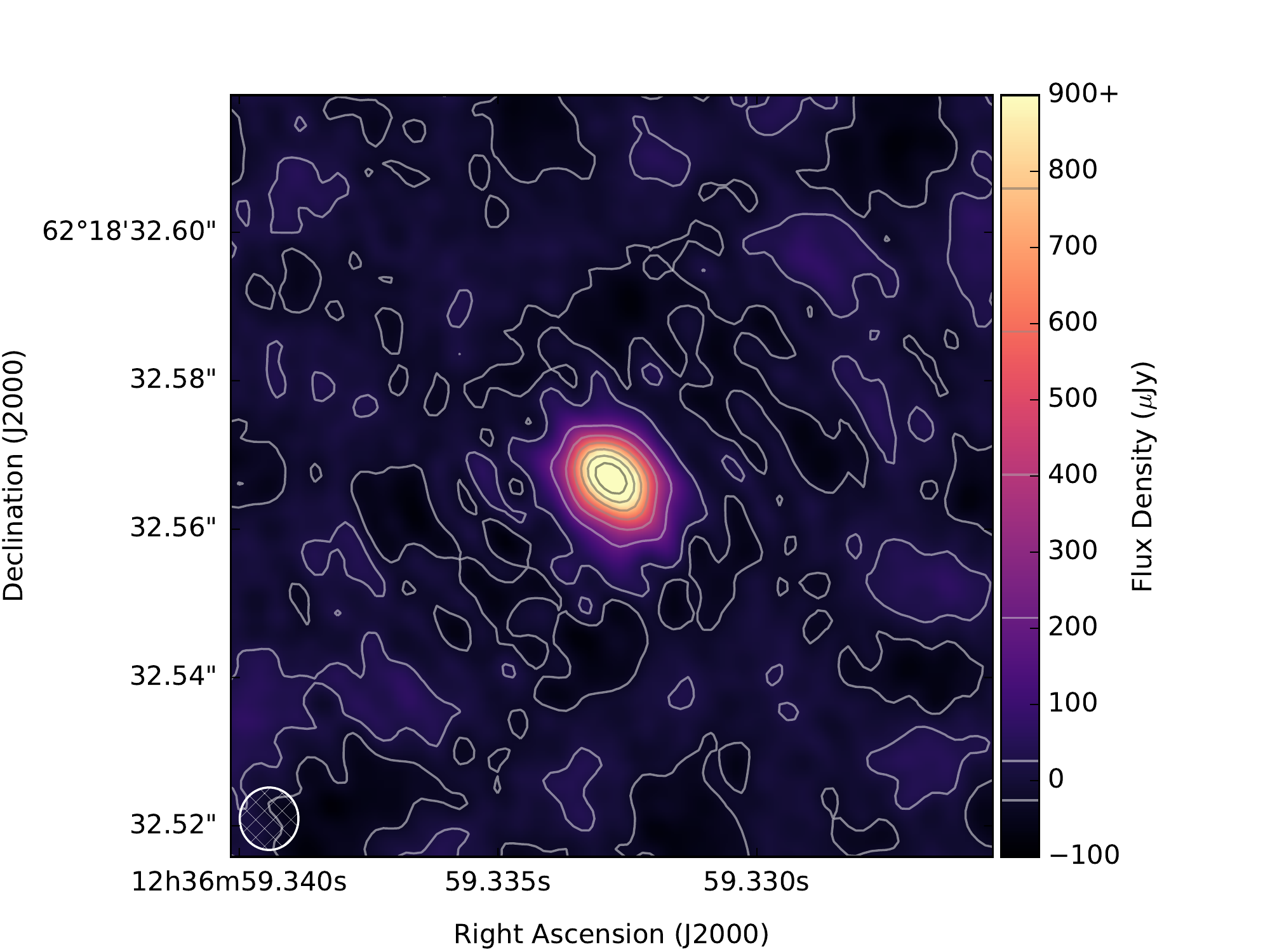}
                \label{fig:sub3}
        \end{subfigure}
        \caption{Compact radio source J123659+621833, which illustrates the effect MSSC has on the structure, fluxes, and noise levels achieved. The colour scale is fixed to the scale of the phase-referenced images to highlight changes in peak brightness and the noise profiles. Contours start at the noise level and are evenly spaced to the peak brightness of each image. \textit{Left panel:} Source when calibrated with only standard phase referencing. This has a peak brightness of 824 $\mu$Jy/beam, integrated flux density of 1.35 mJy and an r.m.s. noise level of 14.7 $\mu$Jy/beam. This gives a maximum S/N of 56.1. Note that the image suffers from significant side lobe negatives next to the source. \textit{Right panel:} The source with MSSC applied. The peak brightness is now 1.28 mJy/beam, integrated flux density of 1.73 mJy and an r.m.s. noise level of 11.1 $\mu$Jy/beam. This results in a greatly improved S/N of 115.8. The side lobe structure has reduced in amplitude and the source is more compact.}
        \label{fig:HDFC0214}
\end{figure*}

The fields were searched for emission using a 6$\sigma$ detection threshold. Figure~\ref{fig:noise_distribution} shows the pixel brightness distribution for a blank field and a field with a source detected. The figure shows that the noise in both fields exhibit non-Gaussianity towards the extrema of the flux distribution. This suggests that there may be some residual RFI in these data. However, the non-Gaussianity appears to be correlated with the flux density of the source. This suggests that residual gain and phase errors are present in these data, which  scatters some of the flux density from the bright source detection into the side lobes. The excess of pixels at the extrema of the flux distribution results in the detection threshold being placed at 6$\sigma$.
 
Nine sources were used in MSSC. These were sources that were detected when imaged with both uniform and natural weighting. If a source was detected in both images, it is highly suggestive that the source can be detected on all baselines. Sources outside the primary beam of the largest telescopes were also avoided due to a-projection effects. The a-projection arises as a result of an intrinsic optical path difference of the radio waves across the primary beam of a telescope, along with time-varying gains that are caused by antenna pointing errors and rotation of asymmetric antenna power patterns \citep[see][]{rau2009advances}.  The MSSC procedure largely follows the techniques outlined in \citet{2013A&A...551A..97M} and is described below. 

Each data set was re-imaged with uniform weighting and was de-convolved with the synthesised beam using the CLEAN algorithm \citep{clark1980efficient}. Each set of visibilities were divided by the CLEAN model using the AIPS task UVSUB. This produced a point source with normalised amplitude, located in the centre of the target field (see Figure \ref{fig:point_plots}, left panel). We note that there was a small reduction in S/N when this step was undertaken. The CLEAN model cannot fully characterise the source structure and, as a result, some of the flux density is scattered into the side lobes. This is shown in the left panel of Figure~\ref{fig:point_plots}. Any offsets in the location of the peak brightness compared to the centroid of the phase centre were removed when the CLEAN model was divided through. UVSUB adjusted the weights ($w_i$) of each data set ($i$) by the inverse square of the amplitude adjustment such that
\begin{equation}
w_i = \left(\frac{1\ \textrm{Jy}}{A_{i}\ \textrm{Jy}}\right)^{2}. \nonumber
\end{equation}

\begin{table*}[ht]
\caption{Comparison between standard phase referencing, single source self-calibration and MSSC calibration techniques.}
        \begin{center}
        \resizebox{\textwidth}{!}{
\begin{tabular}{|l|llll|llll|llll|}
        \multicolumn{1}{c}{} &
        \multicolumn{4}{l}{Phase referencing} &
        \multicolumn{4}{l}{Single source self-calibration} &
        \multicolumn{4}{l}{MSSC} \\
        \hline
        Source ID & P ($\mu$Jy/bm) & I ($\mu$Jy) & N ($\mu$Jy/bm) & S/N & P ($\mu$Jy/bm) & I ($\mu$Jy) & N ($\mu$Jy/bm) & S/N & P ($\mu$Jy/bm) & I ($\mu$Jy) & N ($\mu$Jy/bm) & S/N \\
        \hline
          123608+621036 & \textbf{41.8} & \textbf{50.8} & \textbf{6.3} & \textbf{6.6} & \textbf{45} & \textbf{65.6} & \textbf{6.3} & \textbf{7.1} & \textbf{63.5} & \textbf{78.8} & \textbf{5.6} & \textbf{11.4}\\
  123618+621541 & 61 & 75.2 & 9.7 & 6.3 & 63.7 & 75 & 9.9 & 6.4 & 75.3 & 90.0 & 9.7 & 7.8\\
  123620+620844 & \textbf{60} & \textbf{60} & \textbf{6.9} & \textbf{8.7} & \textbf{37.6} & \textbf{49} & \textbf{6.1} & \textbf{6.2} & 83.4 (\textbf{76.5}) & 83.4 (\textbf{82.1}) & 9.7 (\textbf{6.8}) & 8.6 (\textbf{11.3})\\
  123622+620654 & \textbf{42.8} & \textbf{68.7} & \textbf{6.4} & \textbf{6.7} & \textbf{N-D} & \textbf{N-D} & \textbf{5.91} & - & \textbf{40.8} & \textbf{56.5} & \textbf{5.9} & \textbf{6.9}\\
  123624+621643 & 97.1 & 131.3 & 9.7 & 10.0 & 126.9 & 130.6 & 9.9 & 12.8 & 131.3 & 156.8 & 9.8 & 13.4\\
  123641+621833 & \textbf{37.2} & \textbf{60.8} & \textbf{5.7} & \textbf{6.6} & \textbf{48.3} & \textbf{59.8} & \textbf{5.5} & \textbf{8.8} & \textbf{53.6} & \textbf{58.4} & \textbf{5.8} & \textbf{9.3}\\
  123642+621331 & \textbf{88.2} & \textbf{101.6} & \textbf{5.9} & \textbf{14.8} & 85.9 (\textbf{96.3}) & 106.2 (\textbf{115.5}) & 9.7 (\textbf{6.8}) & 8.9 (\textbf{14.2}) & 72.9 (\textbf{113.2}) & 153.1 (\textbf{141.2}) & 9.6 (\textbf{6.0})  & 7.6 (\textbf{18.9})\\
  123644+621133 & 195 & 197.1 & 9.7 & 20.1 & 155.1 & 168.7 & 9.9 & 15.7 & 262.1 & 256.0 & 9.7 & 27.0\\
  123646+621405 & 76 & 115.2 & 9.6 & 7.9 & 68.7 & 100.5 & 9.6 & 7.2 & 114.1 & 135.0 & 9.5 & 12.0\\
  123653+621444 & \textbf{45.9} & \textbf{53.3} & \textbf{5.3} & \textbf{8.6} & \textbf{49.3} & \textbf{56.2} & \textbf{5.6} & \textbf{8.8} & \textbf{58.3} & \textbf{62.9} & \textbf{5.4} & \textbf{10.8}\\
  123659+621833 & 824 & 1348 & 14.7 & 56.1 & 1284 & 1743.9 & 18.1 & 70.9 & 1284.9 & 1732.7 & 11.1 & 115.8\\
  123700+620910 & \textbf{63} & \textbf{79.1} & \textbf{5.8} & \textbf{10.8} & \textbf{48.8} & \textbf{63.4} & \textbf{5.7} & \textbf{8.5} & 63.4 (\textbf{77.4}) & 89.2 (\textbf{91.1}) & 9.5 (\textbf{5.7}) & 6.7 (\textbf{13.6})\\
  123709+620838 & \textbf{34.5*} & \textbf{41.7*} & \textbf{5.9*} & \textbf{5.8*} & \textbf{28.4*} & \textbf{34.1*} & \textbf{5.4*} & \textbf{5.2*} & \textbf{44.9} & \textbf{55.5} & \textbf{5.5} & \textbf{8.2}\\
  123714+621826 & 170.5 & 181.2 & 10.0 & 17.1 & 185.1 & 216.2 & 10.4 & 17.8 & 235.1 & 252.5 & 10.1 & 23.3\\
  123715+620823 & 840.8 & 946.8 & 13.5 & 62.3 & 524.4 & 701 & 14.6 & 35.9 & 1242.6 & 1300.6 & 15.6 & 79.7\\
  123716+621512 & \textbf{50.1} & \textbf{54.3} & \textbf{5.6} & \textbf{8.9} & \textbf{49.6} & \textbf{59.5} & \textbf{6.1} & \textbf{8.2} & \textbf{59.4} & \textbf{75.1} & \textbf{5.9} & \textbf{10.1}\\
  123717+621733 & 68.3 & 76.2 & 9.7 & 7.0 & 71.8 & 102.8 & 9.7 & 7.4 & 86.3 & 99.3 & 9.6 & 9.0\\
  123721+621130 & 110.7 & 153 & 9.9 & 11.1 & 112 & 143.9 & 9.8 & 11.5 & 182.0 & 195.1 & 9.9 & 18.5\\
  123726+621128 & \textbf{49} & \textbf{63.7} & \textbf{5.6} & \textbf{8.8} & \textbf{N-D} & \textbf{N-D} & \textbf{6.43} & - & \textbf{57.3} & \textbf{66.5} & \textbf{5.6} & \textbf{10.2}\\
  123701+622109 & \textbf{55.6} & \textbf{80.4} & \textbf{5.6} & \textbf{10.0} & \textbf{57.7} & \textbf{73.1} & \textbf{5.7} & \textbf{10.1} & \textbf{64.0} & \textbf{77.8} & \textbf{5.6} & \textbf{11.4}\\
        \hline\end{tabular}
}
\end{center}
\tiny
\begin{tabular}{|l|l|l|l|}
\hline
&Phase referencing & SSSC & MSSC \\
\hline
Detected with natural weighting & 19 & 17 & 20 \\
Detected with uniform weighting & 9 & 10 & 12 \\ 
Failed solutions (Iteration 3) & - & 7\% & 2\% \\
\hline
\end{tabular}
\tablefoot{\textit{Top panel:} Comparison of the peak flux density per beam or brightness (P) in $\mu$Jy/beam (shortened to $\mu$Jy/bm), integrated flux density (I), in $\mu$Jy, r.m.s. noise (R), in $\mu$Jy/beam, and the S/N of the peak brightness to the r.m.s. noise for three different calibration methods. The peak flux densities and integrated flux densities were determined using the AIPS task JMFIT and the noise was measured using the AIPS task IMSTAT. The phase referencing uses only the two designated calibrators J1241+602 and J1234+619. Single source self-calibration has an additional calibration step. Only the brightest detected source (J123659+621833) is used for self-calibration with a solution interval of 6 minutes and MSSC uses 9 sources in MSSC. All entries correspond to values with calibrated weights, apart from entries in bold which are detections with natural weighting. The * represents sources which did not reach the detection threshold of 6$\sigma$ but their flux densities could be measured, whereas N-D (non-detection) indicates sources that did not reach the detection threshold and their flux densities could not be measured. \textit{Bottom panel:} A summary of the total number sources that reached the 6$\sigma$ detection threshold with each calibration technique along with the percentage of failed solutions during the last iteration of self-calibration.} 
        \label{tab:multicol}
        \end{table*}
For example, a source with S/N of 5 will only contribute 1/100 of the combined signal relative to a source with S/N of 50. This effectively maximises the signal to noise when the data sets are combined.

The source coordinates in each data set were changed to the centre of the primary beam and the data sets were concatenated into one set using the AIPS task DBAPP. The choice of source coordinates is arbitrary. All of the source positions were changed to the same coordinates so they could be stacked effectively. This resulted in a data set with visibilities that represent a normalised point source. Each baseline, time, and frequency stamp now contains multiple measurements of a normalised point source. The combination of all detected sources increases the S/N and makes self-calibration possible (see Figure~\ref{fig:point_plots}, right panel).

The visibilities were then self-calibrated in phase using the task CALIB. A normalised point source was used as a model for just the first iteration of self-calibration. These corrections were then applied to the point source data set. The combined data set was imaged with the phase corrections applied and the subsequent image was then used as a model for the next round of self-calibration. This was iterated until the phase corrections converged on zero. To get enough S/N, there is the option of combining spectral windows, polarisations, or increasing the solution interval. All of these can be changed depending on the flux-density distribution in the target field and the sensitivity of the observations. Phase corrections derived were written to AIPS SN tables, which were attached to a dummy UV file using the task FITTP.  These solution (SN) tables were then copied and applied in AIPS to all of the other phase-referenced data sets. This process can be repeated if necessary.

\section{Results}
\label{Sec:Results}
In our example application, phase corrections derived using MSSC were applied to the HDF-N data set. Three self-calibration iterations were conducted using a two minute solution interval. All spectral windows and polarisations were combined. The resulting visibilities were imaged as before with both natural and uniform weighting. The sources detected with MSSC were compared to a standard phase-referenced data set (as described in Section~\ref{sec:2}) and a data set with an additional single source self-calibration applied. The source chosen for this was J123658+621833, which has an integrated flux density of 1.4 mJy. The single source self-calibration comprised of three iterations with a solution interval of 6 minutes. A solution interval was selected which was long enough to provide sufficient S/N for accurate phase-solution determinations and acceptably low solution failure rates, whilst being short enough to correct for the residual phase errors still left in the data set after initial phase-referencing to a nearby calibrator.

For example, using MSSC the combined S/N of the point source was 93, which is lower than the theoretical S/N of 113. This is most likely caused by the inaccuracies in the CLEAN model that was used to characterise the source structure. As a result, some flux density is scattered into the side lobes when the visibilities are divided by the CLEAN model. 

We adopt a solution acceptance threshold of 5$\sigma$ i to reduce scatter in the solutions. The S/N of the combined point source was scaled to the solution interval of 2 mins, which resulted in a S/N of $\sim$3.2. This means that the majority of solutions do not get rejected by our acceptance threshold and, as a result, the number of failed solutions is only 2\%. For example, with a 1-minute solution interval we acquire a scaled S/N of 2.3 and there is much higher scatter, which results in 25\% failed solutions. With higher solution intervals, we found that solution failures rates remained constant but the peak brightness of the target sources decreased, hence a 2-minute interval was found to be optimal. This argument was also used to set the solution interval for single source self-calibration.

When MSSC was compared to standard phase referencing, it was found that all sources exhibited an increase in S/N. On average, the S/N increase was found to be 27\% in naturally weighted images and 63\% in the uniformly weighted images. Twelve sources can be imaged with uniform weighting compared to nine with standard phase referencing. MSSC enabled one more source to reach the detection threshold set at 6$\sigma$. MSSC corrections also provide an improvement in the dynamic ranges and the noise profiles of the images. Figure~\ref{fig:HDFC0214} illustrates this by comparing the phase-referenced set of J123658+621833 to the corresponding MSSC-calibrated data. 

In the single source self-calibrated data, ten sources can be imaged with uniform weighting. When compared to MSSC, we see an average increase in S/N of 36\% in the naturally weighted image and 69\% in the uniformly weighted images. Three sources do not reach the detection threshold in either natural or uniform weighting schemes. We note that single source self-calibration performs worse than standard phase referencing. Even with a solution interval of six minutes, 7\% of all phase solutions fail and, therefore, there are not enough good solutions to improve the image. Whereas with MSSC, we can reduce the solution interval to two minutes with only a 2\% failure rate and gain corrections that dramatically improve the S/N. Table~\ref{tab:multicol} summarises the results presented. 

\begin{figure*}[!htb]
        \centering
        \resizebox{\hsize}{!}{\includegraphics{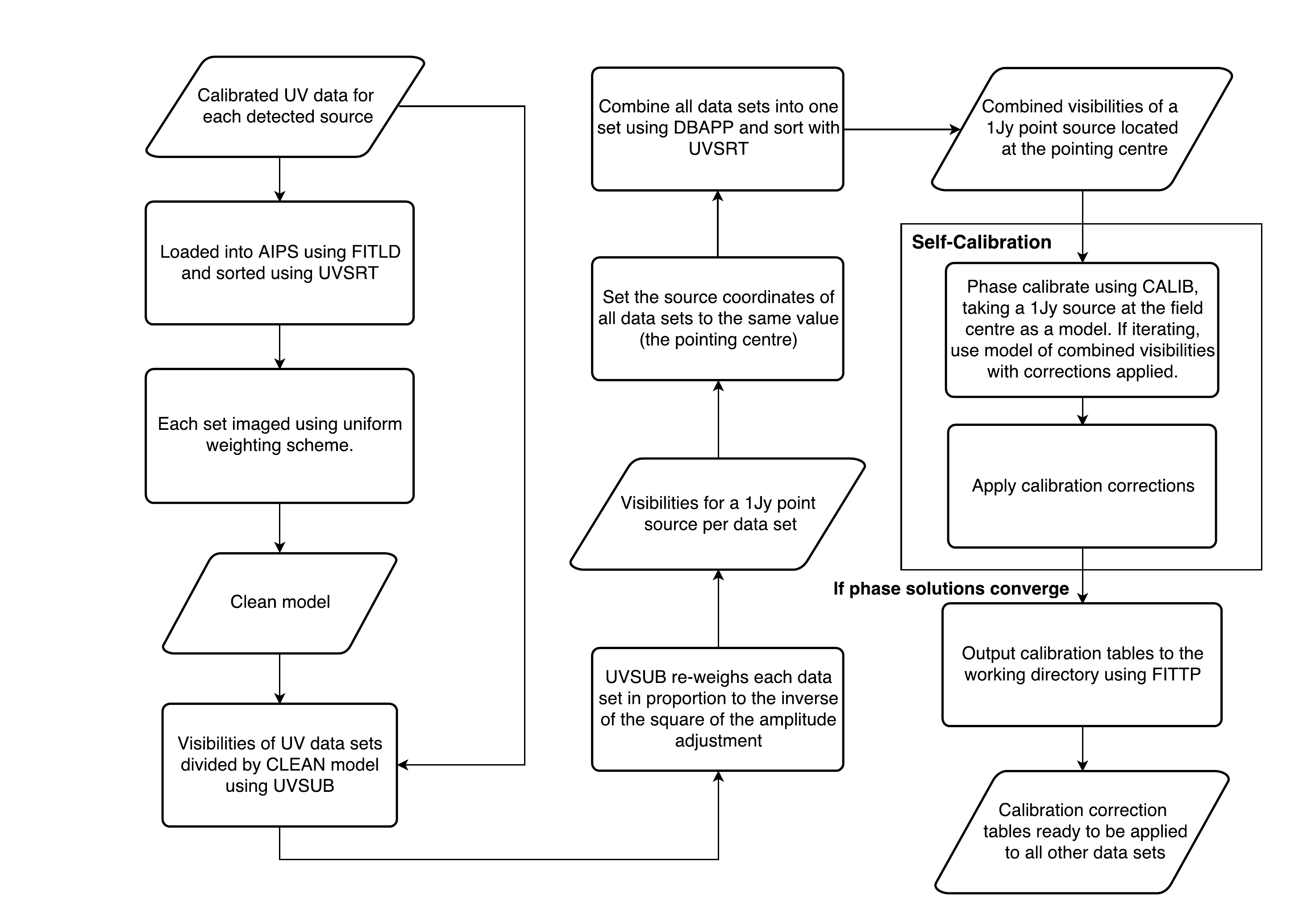}}
        \caption{MSSC algorithm illustrating the various AIPS tasks used to perform the calibration. The final result is the AIPS solution (SN) tables that contain the phase corrections, which can then be applied to the data.}
        \label{Fig:1}
\end{figure*}

\section{Future applications}

With wide-field VLBI becoming more accessible, MSSC can be used as a direction-dependent calibration tool. Direction-dependent calibration techniques are designed to account for atmospheric inhomogeneities and primary beam variations across the field of view. These have been used extensively at lower frequencies where data are particularly susceptible to errors from ionospheric variations. LOFAR, for example, uses the algorithm \texttt{SAGECal} which has been used to great effect to reduce errors from both the ionospheric variations and beam variations \citep[see][]{LOFARNCP2013}. MSSC can also be used as a direction-dependent algorithm by means of faceting. We note that this is different to methods like \texttt{SAGECal} but the intention is the same. In MSSC, phase solutions are essentially the average of the corrections derived for each target source weighted by the square of the brightness of each source. If the target field is split into facets, each of which is an isoplanatic patch, we can separate the sky into subsets of sources corresponding to different areas. MSSC can then be run on each subset of source, which will provide different corrections for each area of the sky.

Using the HDF-N observations, we can approximate the minimum S/N required for MSSC in a typical EVN observation. The minimum S/N can be used to derive the optimal solution interval as described in Section~\ref{Sec:Results}. The observations have a theoretical combined S/N of 113.1 over the duration of the observation, which can be more usefully expressed as the S/N ratio required for a suitable solution interval.
We can scale the combined S/N by the square root of the ratio between the solution interval (2 mins) to the time on the target field (18 hours) to acquire a theoretical minimum array S/N of 3.2. This is, of course, dependent on the number of telescopes and the sources in your field. However it can be used as a guide for deciding if MSSC can be used for the EVN. The combined S/N for any typical EVN observation can be determined by adding, in quadrature, the peak brightness of all target sources within a target field or facet divided by the r.m.s. noise. This can be scaled by the square root of the ratio of the solution interval to the total observing time to acquire the S/N per solution interval. If this value is larger than the minimum value of 3.2, derived from the HDF-N data, then MSSC should perform adequately. Estimating the minimum angular area needed to use MSSC is extremely inaccurate owing to the spatial variability of compact sources and the poorly constrained sub-mJy VLBI flux-density distribution. However, if the sources in the target field far exceed this required S/N, then the sources can be split up into facets containing subsets of sources for which direction-dependent corrections can be derived using MSSC. Using these HDF-N observations as an example of a field with relatively few bright radio sources, we suggest that observations that cover areas greater than 200 arcminute$^2$ would allow direction-dependent solutions to be obtained using MSSC by means of faceting.

In principle, MSSC can be used for \textit{any} future VLBI observation. The growing number of catalogues of mJy sources, most notably from the mJIVE survey \citep{deller2014mjive}, mean that VLBI detected sources are close to almost every target field. By targeting these sources, in multiple simultaneous phase centre observing mode, MSSC can be used on their combined response to improve calibration corrections. MSSC can prove to be an extremely powerful tool to improve the dynamic range of any future VLBI data set.

\section{Conclusions}

We present a new calibration technique termed `multi-source self-calibration' which can be used on wide-field VLBI data sets to increase the phase stability of the target sources. This technique combines in-beam sources to permit phase self-calibration of the target field. It can be used to improve the traditional phase-referencing techniques used in VLBI observations. The multi-field self-calibration algorithm is outlined in Fig.~\ref{Fig:1}. A Parseltongue script is available at \url{https://github.com/jradcliffe5/multi\_self\_cal}, which could be used on any wide-field or future VLBI data set after standard calibration has been applied. The script includes options to change various parameters such as the number of self-calibration iterations and will be revised constantly in the future. 

The MSSC technique is designed for observations of specific faint sources such as GRBs, supernovae remnants, and low-luminosity AGN. In this paper we demonstrated the power of multi-field self-calibration on a 1.6GHz wide-field VLBI observation of the HDF-N. With just standard phase referencing, 20 sources were detected but many of the images were limited in dynamic range or only detected using natural weighting. When applied, the technique significantly improved the S/N of all sources imaged (see Section \ref{fig:HDFC0214}) and allowed three more sources to be detected when imaged with uniform weighting.

{With rapidly improving sensitivities and correlator capabilities of VLBI arrays, observations of multiple primary beams is now possible. MSSC  permits VLBI observations to be conducted in any direction on the sky and can allow directional-dependent calibration to be performed. New instruments, such as the e-EVN and possibly VLBI with the upcoming SKA, will make $\mu$Jy source detection on VLBI baselines routine and, as such, increase the wealth of potential calibrators that can be used in multi-source self-calibration. MSSC could prove to be a very powerful tool in unveiling the microJy regime of compact radio sources. 

\begin{acknowledgements}
The research leading to these results has received funding from the European Commission Seventh Framework Programme (FP/2007-2013) under grant agreement No 283393 (RadioNet3).\\

The European VLBI Network is a joint facility of European, Chinese, South African, and other radio astronomy institutes funded by their national research councils.

We'd like to thank the anonymous referee for their helpful comments. 
\end{acknowledgements}

\bibliographystyle{aa}
\bibliography{Multi_field_self_calibration}

\end{document}